\documentclass[prl,aps,twocolumn,showpacs]{revtex4}
\usepackage{graphicx}
\begin{document}

\title{Electron interaction with domain walls in antiferromagnetically coupled
multilayers}

\author{F. G.~Aliev$^{1,\, *}$, R.~Schad$^2$, A.~Volodin$^3$, K.~Temst$^3$,
C.~Van~Haesendonck$^3$, Y.~Bruynseraede$^3$, I.~Vavra$^4,$ V. K.~Dugaev$^{5,6}$,
and R.~Villar$^1$}

\affiliation{$^1$Dpto. Fisica Materia Condensada, C-III, Universidad Autonoma de Madrid, Spain}
\affiliation{$^2$CMIT, University of Alabama, Tuscaloosa, AL 35487, USA}
\affiliation{$^3$Laboratorium voor vaste Stofffysica en Magnetisme, K. U. Leuven, B-3001 Leuven, Belgium}
\affiliation{$^4$Institute of Electrical Engineering - SAS, 84239 Bratislava, Slovakia}
\affiliation{$^5$Max-Planck-Institut f\"ur Mikrostrukturphysik, Weinberg 2, 06120 Halle, Germany}
\affiliation{$^6$Institute for Problems of Materials Science,
National Academy of Sciences of Ukraine, 58001 Chernovtsy, Ukraine}

\date{\today}

\begin{abstract}
For antiferromagnetically coupled Fe/Cr multilayers the low field
contribution to the resistivity $\rho _{DW}$, which is caused by
the domain walls, is strongly enhanced at low temperatures. The
low temperature resistivity $\rho _{DW}$ varies according to a
power law $\rho _{DW}(T)=$ $\rho _{DW}(0)-A\; T^\alpha $ with
the exponent $\alpha \simeq 0.7 - 1.$ This behavior can not be
explained assuming ballistic electron transport through the
domain walls. It is necessary to invoke the suppression of
anti-localization effects (positive quantum correction to conductivity)
by the nonuniform gauge fields caused by the domain walls.
\end{abstract}


\maketitle

Renewed interest in the domain wall (DW) contribution to the resistivity is
stimulated by its relevance for fundamental physics
\cite{Berg78,Gio95,Gregg96,Levy97,Ru98} and possible applications. Indeed, domain
walls may strongly influence the electrical noise and operation of
magnetoelectronics devices \cite{Gider98}. Although the number of DWs was
controlled and directly observed in Fe \cite{Ru98} and in Co films \cite
{Gregg96} at room temperature, where DW formation is relatively well
understood, no clear picture has emerged allowing to explain the results.
The anisotropic magnetoresistance (AMR) dominates the low field
magnetoresistance and complicates the extraction of the true DW contribution
to the resistivity \cite{Rued99}. In order to minimize the AMR contribution,
thin films with reduced magnetization and special DW configuration have been
studied \cite{Viret2000}. Apart from the ballistic contribution to the DW
magnetoresistance \cite{VanH99}, quantum interference also affects the
electron transport through DWs \cite{Tatara97,Geller98}.

Antiferromagnetically (AF) coupled magnetic multilayers (MMLs) are systems
with reduced magnetization and consequently a strongly suppressed AMR. At
high temperatures, weak pinning of the DWs in the MMLs is expected to
suppress the DW magnetoresistance. For fixed magnetic field the DW
magnetoresistance should emerge at sufficiently low temperatures where DWs
become strongly pinned and their configuration is not affected by thermal
fluctuations or by the applied electrical current.

Here, we report on our detailed study of the low field electrical
resistivity in AF coupled Fe/Cr MMLs. The well known giant magnetoresistance
(GMR) in this system is dominated by a realignment of the magnetization
direction in adjacent magnetic layers \cite{Fert95}. The presence of DWs
should result in an additional, small in-plane magnetoresistance \cite
{Zhang94}. While the GMR is known to saturate at low temperatures \cite
{Mattson91}, the temperature dependence of DW magnetoresistance is still a
matter of controversy. In order to separate the two contributions, we
performed a systematic study of the temperature dependence of the
resistivity in low magnetic fields. Our main findings are that (i) the
presence of DWs in an AF coupled MML does not affect the resistivity at room
temperature, and (ii) at low temperatures the DW contribution to the
resistivity becomes positive and strongly temperature dependent. We explain
these observations in terms of the suppression of positive quantum
correction to conductivity (so called ''anti-localization'' effect) by the
domain walls.

Epitaxial [Fe/Cr]$_{10}$ multilayers with 10 bilayers are prepared in a
molecular beam epitaxy system on MgO (100) oriented substrates held at $%
50\,^{\circ }{\rm C}$ and covered with an approximately $10\,{\rm \AA }$
thick Cr layer. The thickness of the Fe layer was varied between 9 and $30\,%
{\rm \AA }$, while the Cr layer thickness (typically 12 to 13 \AA )
corresponds to the first antiferromagnetic peak in the interlayer exchange
coupling for the Fe/Cr system \cite{Parkin93} and produces a maximum GMR
which is about $20\,\%$ at $300\,{\rm K}$ and $100\,\%$ at $4.2\,{\rm K}$. A
commercial cryogenic system (PPMS, Quantum Design) was used to measure
magnetization, magnetic susceptibility, and electrical resistance with a
standard four-probe ac method at a frequency of $321\,{\rm Hz}$ with
currents ranging between 15 and $50\,\mu {\rm A}$. The magnetic fields
created by these currents are well below $0.1\,{\rm Oe}$ and do not affect
the DWs. The magnetic field dependence of the susceptibility along different
crystallographic directions as well as the low residual resistivity
(typically less than $13\,\mu \Omega \, {\rm cm}$ at a saturation field of $1\,%
{\rm T}$) confirm the good epitaxial growth of our MMLs. Magnetization
measurements at $4.2\,{\rm K}$ reveal that the antiferromagnetic fraction $%
(1-M_r/M_s)$, with $M_r$ and $M_s$ the remanent and the saturation
magnetization, respectively, exceeds $80\,\%$. This indicates that bilinear
AF coupling dominates over biquadratic exchange coupling \cite{BC}. The
existence of a small non-compensated magnetic moment may allow DW motion in
our artificial antiferromagnet.

\begin{figure}
\vspace{-1cm}
\includegraphics[scale=0.5]{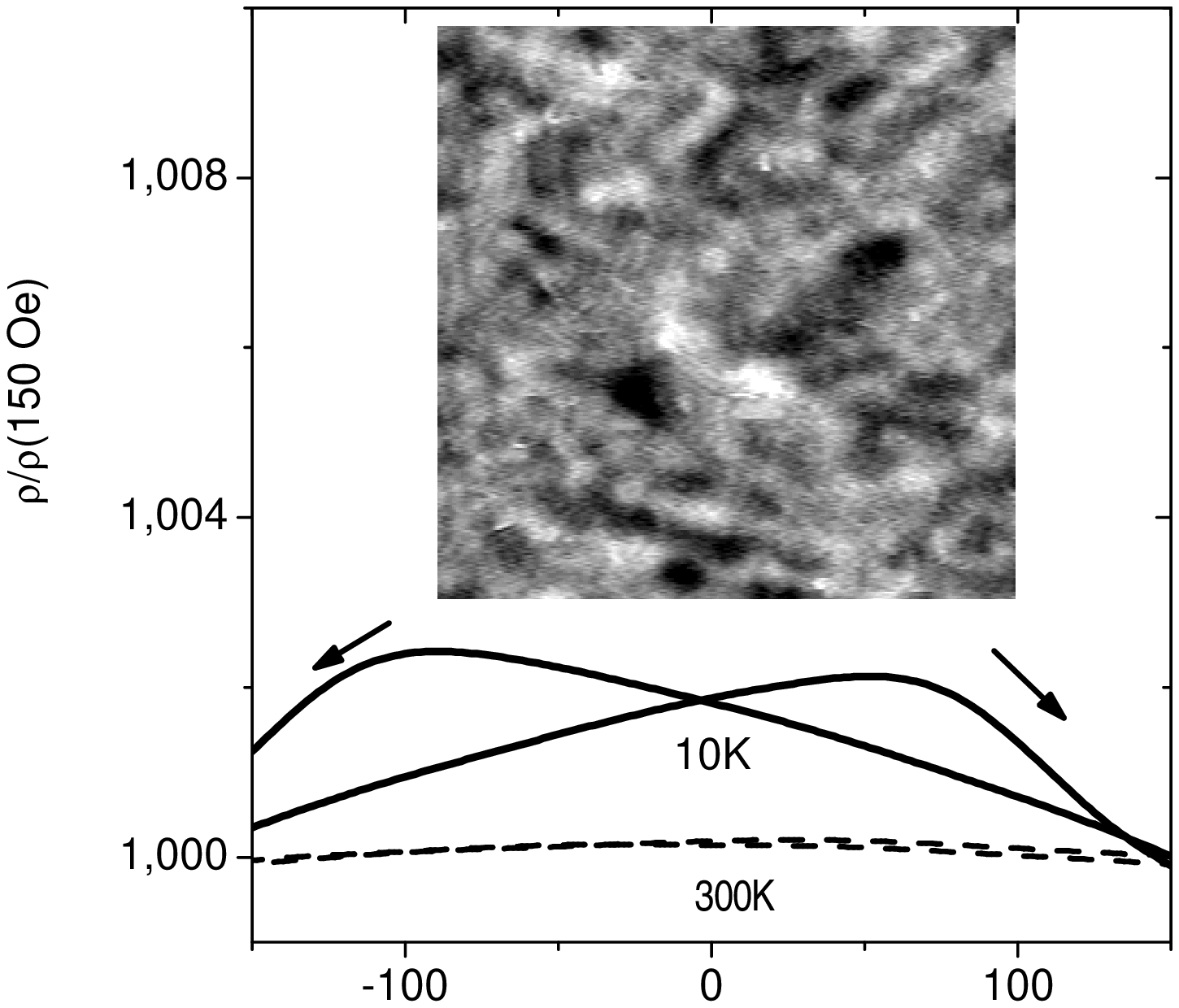}
\vspace{-6.5cm}
\caption{Room temperature and low temperature normalized magnetoresistance $%
\rho (H)/\rho (150\,{\rm Oe})$ for an [Fe($12\,{\rm \AA }$)/Cr]$_{10}$
multilayer with the current $I$ parallel to the field $H$ and parallel to
the (110) direction. The inset shows a typical MFM image ($8\times 8\,\mu
{\rm m}^2$) of AF coupled [Fe/Cr]$_{10}$\ multilayer at $4.2\,{\rm K}$}
\end{figure}

The inset in Fig.~1a shows a typical magnetic image of an AF coupled Fe/Cr
MML at $T=4.2\,{\rm K}$ (image size is $8\times 8\,\mu {\rm m}^2$) using a
home-built cryogenic magnetic force microscope (MFM) \cite{Volodin}. The MFM
picture, which ''feels'' magnetic contrast, reveals different irregularly
shaped domain walls (which is a characteristic feature of strong AF coupling
\cite{Fert95,Ruhrig91}) with micrometer dimensions. While our MFM
measurements reveal a similar domain structure at room temperatures, the
magnetoresistance curves, which are shown in Fig.1, are very different. The
low field magnetoresistance is strongly enhanced at low temperatures. The
susceptibility data point towards a weak pinning of the DWs at room
temperature and a strong pinning at low temperatures \cite{Aliev02}.

\begin{figure}
\vspace{-2.5cm}
\includegraphics[scale=0.56]{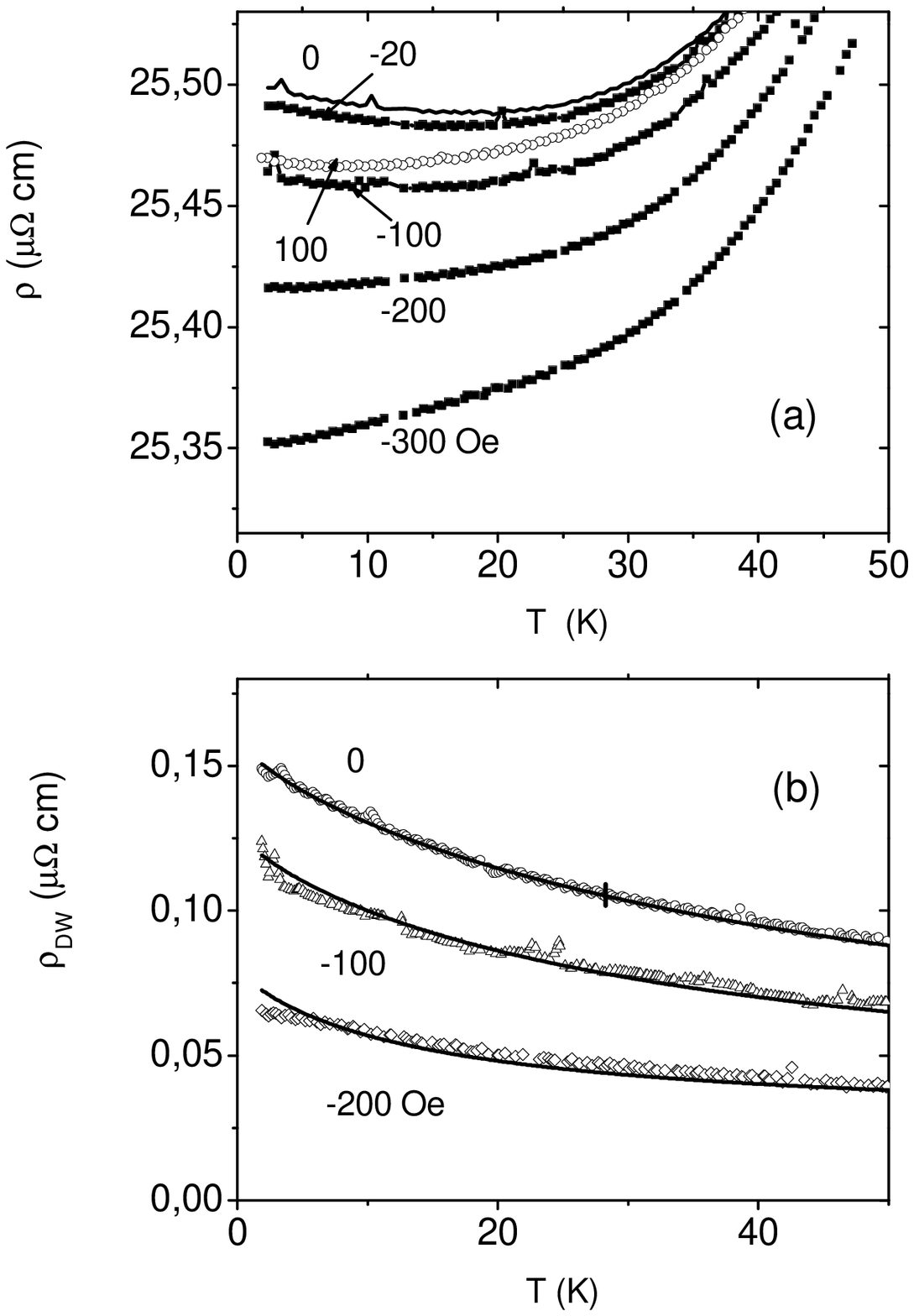}
\vspace{-3cm}
\caption{(a)\ Temperature dependence of the resistivity for [Fe($12\,{\rm \AA }$)/Cr]$%
_{10}$ multilayer in different magnetic fields. Both the field and the
current are along the (110) axis. (b) Temperature dependence of the DW
contribution to the resistivity. $\rho _{DW}=\rho (T,H)-\rho (T,H_S)$
determined from the data shown in (a) for $H_S=-300\,{\rm Oe}$. The solid
lines correspond to the fits which are described in the text.}
\end{figure}

Figure~2a shows the temperature dependence of the electrical resistivity
$\rho $ for an [Fe($12\,{\rm \AA }$)/Cr]$_{10}$ MML for different magnetic
fields ($\left| H\right| \leq 300\,{\rm Oe)}$. The magnetic field is applied
in the plane of the film and is parallel to the current as well as to the
longer side of the rectangular ($5\times 25\,\mu {\rm m}^2$) sample which is
directed along the (110) axis. For $\left| H\right| >100\,{\rm Oe}$ the
$\rho (T)$ dependence reveals a metallic behavior, while for $\left| H\right|
\leq 100\,{\rm Oe}$ there appears a shallow minimum in the $\rho (T)$
curves. We note that in the $\rho (T)$ curves measured after crossing zero
field there appear aperiodic peaks when the applied current is smaller than $%
20\,\mu {\rm A}$, which correspond to an intrinsic noise process in the
sample. The peaks, which can be linked to Barkhausen noise, gradually
disappear when doubling the electrical current or when the absolute value of
the magnetic field exceeds $300\,{\rm Oe}$ \cite{Hardner95}.

A straightforward way to determine the magnetoresistivity of the DWs is to
subtract the temperature dependences of the resistivity measured in the
presence and in the absence of DWs, respectively. However, the magnetic
field $H_S^0$ which guarantees nearly uniform N\'eel vector along the
external field (according to our magnetic susceptibility data $300\,{\rm Oe}%
<\left| H_S^0\right| <1000\,{\rm Oe}$), not only sweeps the DWs out of the
sample, but may also change the angle of the magnetization between adjacent
magnetic layers from the antiparallel alignment (GMR). In order to separate
the magnetoresistivity induced by the GMR effect from the magnetoresistivity
induced by the DWs, we define $\rho _{DW}=\rho (T,H)-\rho (T,H_S)$ with $%
\left| H_S\right| \leq 300\,{\rm Oe}$. Although this method may
underestimate the magnetoresistivity of the DWs because not all domains will
be removed by the applied field $H_S$, the method provides a possibility to
determine the temperature dependence of the DW magnetoresistivity. In
Fig.~2b we show $\rho _{DW}(T,H)$ between 1.9 and $100\,{\rm K}$ for
different magnetic fields ranging between $-200\,{\rm Oe}$ and zero field
for $H_S=-300\,{\rm Oe}$. We find that, in contrast to the GMR, the DW
magnetoresistivity is strongly temperature dependent with no sign of
saturation at low temperatures.

Assuming that the magnetic field mainly changes the effective DW
concentration $n_{DW}$ \cite{DWconcentration}, we expect $\rho _{DW}$ to
scale according to $\Delta \rho _{DW}=\rho _{DW}(0)-\rho _{DW}(T)\propto
n_{DW}(H)\; \rho _{DW}^0(T)$ with $\rho _{DW}^0(T)$ a function describing
the temperature dependent electron interaction with DWs. Determined in this
way $\rho _{DW}^0(T)$ which is{\bf \ }independent of the choice of $H_S$ as
long as $\left| H_S\right| \leq 300\,{\rm Oe}$. Our data analysis reveals
that the DW resistivity is roughly given by $\Delta \rho _{DW}\sim
n_{DW}(H)\cdot T^{0.7}$ (Fig.~3a) illustrates the scaling $\Delta \rho
_{DW}\propto T^{0.7}$ for different magnetic fields $\left| H\right|
<H_S=-300\,{\rm Oe}$ for temperatures between $1.9\,{\rm K}$ and $25\,{\rm K}
$. For comparison we also show the qualitatively different temperature
scaling for the GMR (see dashed line in Fig.~3a). The vertical bar in Fig.~2b
estimates the maximum influence of the GMR effect on our data. This
estimation was obtained from the temperature dependence of the
magnetoresistivity measured for two different magnetic fields sufficiently
large to remove all DWs. In agreement with previous results \cite
{Mattson91,Aliev98}, both saturation field and GMR are weakly temperature
dependent below 50K, GMR saturates as $T^2$ and changes in less than 7\%.

\begin{figure}
\vspace{-1cm}
\hspace{-2cm}{\includegraphics[scale=0.53]{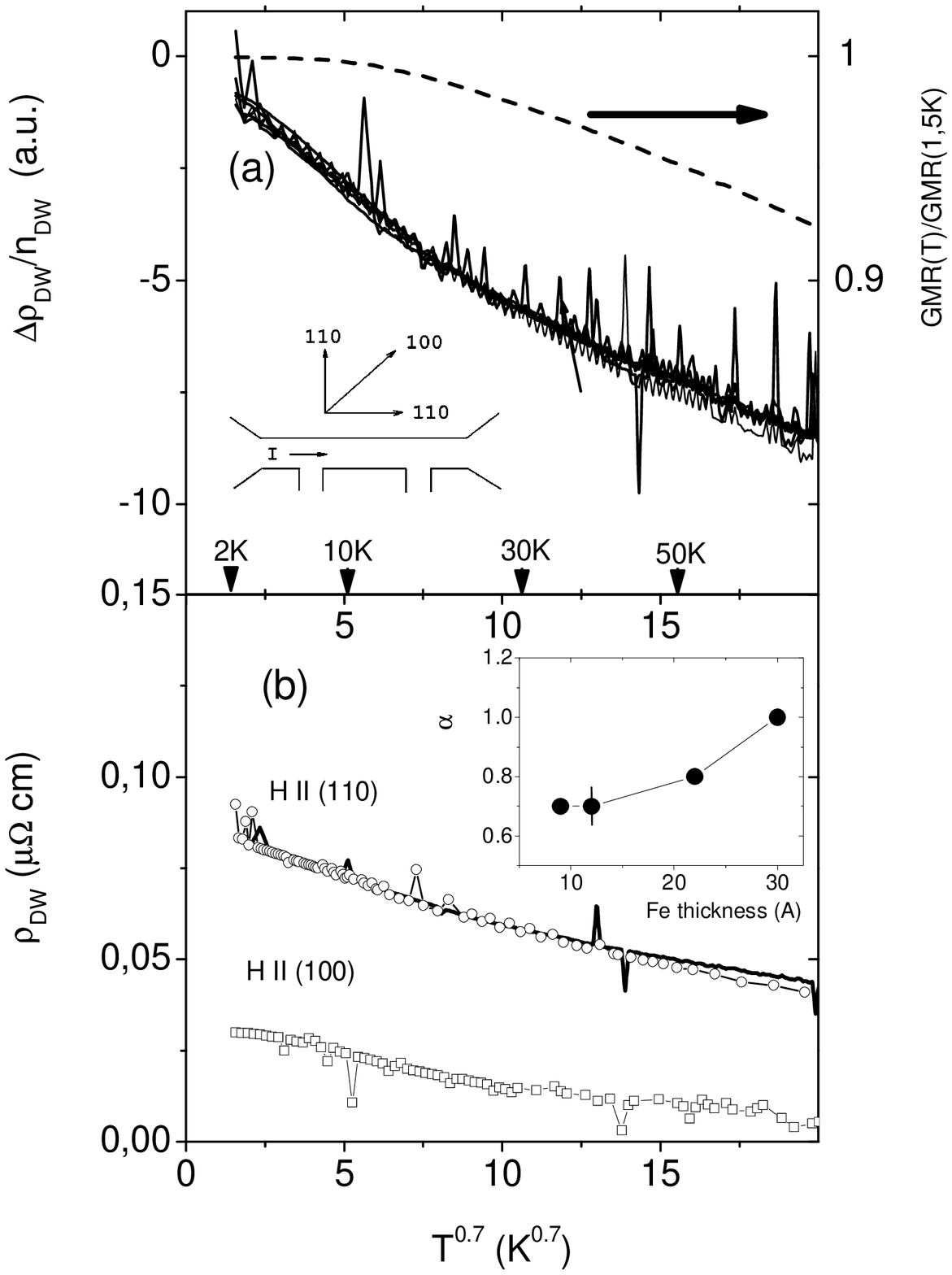}}
\vspace{-3cm}
\caption{(a) Normalized temperature dependence of the DW contribution to the
resistivity. The data have been obtained for magnetic fields: $%
+100,+20,0,-20,-100$ and $-200\,{\rm Oe}$. ${n_{DW}(H)}$ is the
concentration of domain walls, and the DW resistivity has been determined
for $H_S=-300\,{\rm Oe}$. The dashed line corresponds to the GMR
contribution which is obtained from the $\Delta \rho _{DW}(T)$ dependence
for $H=-500\,{\rm Oe}$ with $H_S=-1000\,{\rm Oe}$. The inset gives a
schematic view of the sample geometry. (b) Temperature dependence of the DW
contribution to the resistivity for an [Fe($12\,{\rm \AA }$)/Cr]$_{10}$
multilayer when $H=0$ and for $H_S=-200\,{\rm Oe}$ directed along the (110)
direction. The circles correspond to the case where the field $H$ is
perpendicular to the current $I$, while the solid line corresponds to the
case where $H$ is parallel to $I$. The open squares give $\rho _{DW}$ for $%
H=0$, but with $H_S=-200\,{\rm Oe}$ directed along the (100). The inset
illustrates the dependence of the scaling exponent $\alpha $ on the Fe layer
thickness.}
\end{figure}

Next, we demonstrate that neither the AMR, which depends on the relative
orientation of the magnetization and the current $I$, nor the ordinary
magnetoresistivity (caused by the Lorentz force), which depends on the
relative orientation of $I$ and the magnetic induction $B$, contribute to $%
\rho _{DW}$. The upper curves in Fig.~3b correspond to $\Delta \rho
_{DW}(H_S=-200\,{\rm Oe}$ for the current parallel to (line) or
perpendicular to (circles) the magnetic field $H$ applied parallel to the
(110) direction (see inset in Fig.~3a). If the AMR affects the low field
magnetoresistivity, its contribution will be positive when the field is
parallel to the current and negative when the field is perpendicular to the
current \cite{Berg78}. It is, however, clear that $\Delta \rho _{DW}$ is
almost identical for both cases, indicating the AMR effects can be
neglected. The magnetic field dependence of the DW resistivity is reduced
when the field is applied along the (100) axis (see lower curve in Fig.~3b).
This probably reflects the presence of a crystal lattice induced anisotropy
in the potential barrier which pins the DWs. We have obtained similar
results with a slightly different scaling of the low temperature DW
magnetoresistivity for three other AF coupled Fe/Cr samples with an Fe layer
thickness of 9, 22 and $30\,{\rm \AA }$, respectively. The inset in Fig.~3b
shows the dependence of the scaling exponent $\alpha $ on the Fe thickness.%
{\bf \ }This may reflect a change of the exponent $p$ in the temperature
dependence of the phase breaking time $\tau _\varphi \propto T^{-p/2}$\
which should occur between the ''dirty'' (\ $p=3/2$) and ''clean'' ($p=2$)
limits \cite{Lee85}.

A ballistic approach for the electron transport through DWs \cite{VanH99}
requires that the mean free path $\ell $ in our epitaxial layers exceeds the
DW width $D$ with $20<D<200$~nm for Fe/Cr/Fe trilayers \cite{Schmidt99}.
Therefore, the
condition for ballistic transport may only be fulfilled at sufficiently low
temperatures \cite{Schad98}. Although non-ballistic effects have not yet
been incorporated into the theory \cite{VanH99}, we believe that they cannot
account for the strong variation of $\rho _{DW}$ down to $1.9\,{\rm K}$,
because the mean free path is expected to saturate at low temperatures.
Moreover, the strong pinning of DWs at low temperatures \cite{Aliev02}
implies that a distortion of the current lines by domain walls \cite{Berg78}
or a change of the DW configuration cannot account for the {\em strongly
temperature dependent low field contribution to the magnetoresistivity in
antiferromagnetically coupled magnetic multilayers}.

In order to explain the strong variation of the DW magnetoresistivity at low
temperatures, one has to go beyond the classical approach \cite{VanH99}. A
possibility is to link the observed phenomena either to standard, disorder
related, weak localization effects or to scattering by isolated spins. Our
experimental results are in conflict with both scenarios since the
resistivity correction with and without magnetic field is different when
applying the magnetic field along the hard or along the easy axis (see
Fig.~3b). Moreover, we observe that $\rho (T,H)$ is different when the
magnetic field is changed at low temperature ($4.2\,{\rm K}$) or at high
temperature ($T>150\,{\rm K}$). Finally, we observe some asymmetry in the $%
\rho (T,H)$ data taken for fields with the same amplitude but applied along
different directions (see data for $H=100\,{\rm Oe}$ and $H=-100\,{\rm Oe}$
in Fig.~2a).

Both \cite{Tatara97} and \cite{Geller98} predict a destruction of weak
electron localization by the domain walls, although the details of the
destruction mechanism are different. {\em Direct application of these models
results in a sign of the DW magnetoresistivity which is opposite to the sign
of the experimentally observed magnetoresistance}. However, the sign of the
localization correction may be reversed due to strong spin-orbit (SO)
scattering (anti-localization) \cite{Knap96}. The suppression of the weak
localization corrections by a DW, predicted in \cite{Tatara97,Geller98}, is
related to the effective gauge potential created by the domain wall. In
contrast to the electromagnetic vector potential, which can be linked to an
external magnetic field, the gauge field depends on the spin, giving rise to
a different influence of the domain wall on the different components of the
so-called Cooperon \cite{Geller98}. Our measurements are consistent with an
anti-localization effect in the absence of DWs ($H>300\,{\rm Oe}$) which is
suppressed in the presence of DWs ($H=0$). The appearance of
anti-localization is due to the SO scattering which suppresses the triplet
Cooperons and does not affect the singlet Cooperon \cite{Lee85}.

The SO interaction should be more pronounced in the case of multilayered
structures than in single films. The potential steps at the interfaces in
combination with the relativistic terms in the Hamiltonian may produce
strong SO scattering. The corresponding theory for the interface SO
interaction has been proposed by Bychkov and Rashba \cite{Rashba84}. In the
case of Fe/Cr multilayers the potential steps are about 2.5~eV for the
majority electrons, and one can expect a significant SO scattering from the
interface. In case of strong SO scattering the magnetoresistance is caused
by the destruction of the singlet Cooperon by the gauge field of the DWs.
The model \cite{Tatara97} predicts a suppression of all components of the
Cooperon, while the approach of Lyanda-Geller {\em et al.} \cite{Geller98}
relies on the suppression of some of the components.

When $\ell \ll D$, we can characterize the system in terms of a local
conductivity, which is defined as an average over distances larger than $%
\ell $ but smaller than $D$. For the local conductivity inside a DW we can
estimate the localization correction that is determined by smaller diffusive
trajectories with size $L < D$ as well as by large trajectories $D < L <
L_\varphi$. $L_\varphi $ is the phase relaxation length governing the
destruction of the interference effects. The localization corrections
associated with the small trajectories are suppressed by the gauge field
since they are located within the DW. The contribution of large trajectories
to localization is small, and for strong spin-orbit scattering the local
conductivity within a DW is
\begin{equation}
\sigma _{DW}\simeq \sigma _0+\frac{e^2}{4\pi ^2\hbar }\left[ \frac 1\ell%
-\left( \frac 1{L_{DW}^2}+\frac 1{L_\varphi ^2}\right) ^{1/2}\right] \; ,
\label{1}
\end{equation}
where $L_{DW}$ is the characteristic length which is determined by the
influence of the gauge potential ${\bf A}$. Its magnitude can be estimated
as $A \sim 1/D$, and consequently $L_{DW} \sim D$.

Since the anti-localization correction without DW is
\begin{equation}
\sigma \simeq \sigma _0+\frac{e^2}{4\pi ^2\hbar } \left( \frac 1\ell-\frac 1{%
L_\varphi }\right) \;,  \label{2}
\end{equation}
we are able to estimate the difference in magnetoresistivity due to the DWs
as
\begin{equation}
\sigma _{DW}-\sigma \simeq -\frac{e^2}{4\pi ^2\hbar }\left[ \left( \frac 1{%
L_{DW}^2}+\frac 1{L_\varphi ^2}\right) ^{1/2}-\frac 1{L_\varphi }\right]
\label{3}
\end{equation}
by taking into account that $\sigma _{DW}-\sigma \simeq -\Delta \rho
_{DW}\; \sigma ^2.$ The most important feature of our evaluation of the
anti-localization effects is the fact that the correction to the local
conductivity is determined by the gauge field inside the DW. If the current
flow crosses the DWs, the corrections to the local conductivity, calculated
for a narrow region inside a DW, show up in the sample resistance.

We can also estimate the influence of an external magnetic field of $300\,%
{\rm Oe}$ and of the internal magnetization on the localization corrections.
These effects are small when the magnetic length $l_H\equiv \sqrt{\hbar
c/(eH)}\gg \ell $. For $H=300\,{\rm Oe}$, $l_H\simeq 1.4\times 10^{-5}\,{\rm %
cm}$. Assuming the internal magnetic induction $B=2\,{\rm T}$ (typical value
for Fe), we obtain the corresponding length $l_H\simeq 1.8\times 10^{-6}\,%
{\rm cm}$. On the other hand, the mean free path $\ell \simeq 10^{-7}\,{\rm %
cm}$. Thus, both the external magnetic field and the magnetization are
unable to effectively suppress the anti-localization corrections.

We are able to fit our data to Eq.~(3) when we assume that $L_{DW}$ is
independent of temperature and that the phase breaking length $L_\varphi $
varies with temperature according to a power law $L_\varphi \propto T^{-p/2}$
\cite{Lee85}. On the other hand, we have to introduce an additional
(constant) shift of the data which takes into account the change of the
resistance due to the variation of the angle between magnetic layers. It is
important to note that the three different fits presented in Fig.~2b
correspond to the same fitting parameters (with\ $p=3/2$), except for the
parameter which describes the magnetic contrast ($L_{DW}$). We find that the
effective DW width $L_{DW}$ becomes about 2.5 times larger when the magnetic
field is increased from 0 to $200\,{\rm Oe}$.

We thank A.~P.~Levanuyk and V. V. Moshchalkov for discussions. The research was
supported in parts by Spanish MCyT under\ project BFM2000-0016, by Comunidad
de Madrid (07N/0050/2002) and by Polish Grant PBZ/KBN/044/P03/2001 (V.D.).


\begin{thebibliography}{99}
\bibitem{DWconcentration}  From the inset to Fig. 1a it is clear that in our
Fe/Cr samples the electron waves propagate in a strongly non-uniform
magnetic field and the classical interpretation of domain walls may not be
correct in this case. However, we continue to use the standard terminology.
In our samples the scaling parameter $n_{DW}$ (concentration of DWs) rather
reflects the degree of non-uniformity of the local in-plane magnetic field
inside the Fe layers.
\end{thebibliography}
\end{document}